\input{aipcheck}

\documentclass[
    ,final            
  ]
  {aipproc}

\layoutstyle{8x11double}

\begin{document}

\title{Magnesium Isotopes in Halo Stars}


\author{Jorge Mel\'endez}{
  address={Research School of Astronomy \& Astrophysics, Mt. Stromlo Observatory, Cotter Rd., Weston Creek, ACT 2611, Australia}
}

\author{Judith G. Cohen}{
  address={Palomar Observatory, California Institute of Technology, Pasadena, CA 91125}
}

\begin{abstract}
We have determined Mg isotope ratios in halo field dwarfs and giants in the globular cluster M71
based on high S/N high spectral resolution (R = 10$^5$) Keck HIRES spectra.
Unlike previous claims of an important contribution from intermediate-mass
AGB stars to the Galactic halo, we find that our $^{26}$Mg/$^{24}$Mg ratios 
can be explained by massive stars.

\end{abstract}

\maketitle


\section{Introduction}

Magnesium is composed of three stable isotopes $^{24}$Mg, $^{25}$Mg and $^{26}$Mg,
which can be formed in massive stars (e.g. Woosley \& Weaver 1995).
The lightest isotope is formed as a primary isotope from H, 
while $^{25,26}$Mg are formed as secondary isotopes.
The heaviest Mg isotopes are also produced in intermediate-mass 
AGB stars (Karakas \& Lattanzio 2003), 
so the isotopic ratios $^{25,26}$Mg/$^{24}$Mg increase
with the onset of AGB stars.
Therefore, Mg isotopic ratios in halo stars 
could be used to constrain the rise of AGB stars in our Galaxy.

It is important to know when AGB stars begin to enrich the halo
in order to disentangle the contribution of elements produced by
intermediate-mass stars from those produced by massive stars. 
For example, the high nitrogen abundances observed in metal-poor stars 
can be explained by fast-rotating massive stars 
(e.g. Chiappini et al. 2006) 
or alternatively by intermediate-mass stars, although 
the latter option may be unlikely because those stars 
may not have had time to enrich the halo due to their longer lifetime.

The study of Mg isotope ratios is also important to understand 
the abundance variations in globular clusters 
(e.g. Yong, Aoki \& Lambert 2006; Prantzos, Charbonnel \& Iliadis 2007).

Mg isotopic abundances can be obtained from the analysis
of MgH lines in cool stars (e.g. Boesgaard 1968; Tomkin \& Lambert 1980; 
Barbuy 1985, 1987; McWilliam \& Lambert 1988; Gay \& Lambert 2000; 
Yong, Lambert \& Ivans 2003; Mel\'endez \& Cohen 2007).

In this work, we determine Mg isotopic ratios in
cool halo dwarfs and giants in the globular cluster M71
employing high resolution high S/N spectra taken
at Keck.

\section{Mg isotopes in halo dwarfs}

Three cool halo dwarfs were observed at Keck I employing the upgraded HIRES 
(Vogt et al. 1994) which provides now a maximum resolving power of  R $\approx$ 10$^5$. 
For details see Melendez \& Cohen (2007).

The isotopic ratios were obtained from three wavelength 
regions at 5134.6, 5138.7 \& 5140.2 \AA\ 
(e.g. McWilliam \& Lambert 1988; Yong et al. 2003).
Isotope ratios were obtained by spectral synthesis using a
new line list including atomic lines, C$_2$ \& MgH lines (Mel\'endez \& Cohen 2007).
Macroturbulence was estimated using the Ni I 5115.4 \AA\ and Ti I 5145.5 \AA\ lines, 
and other lines around 5585 \AA.

Low $^{25,26}$Mg/Mg ratios were obtained (Mel\'endez \& Cohen 2007), in
agreement with bona-fide halo dwarfs from Yong et al. (2003).
These low ratios are in good agreement with yields of
massive stars (e.g. Goswami \& Prantzos 2000; 
 Fenner et al. 2003).

\begin{figure}
  \includegraphics[height=.25\textheight]{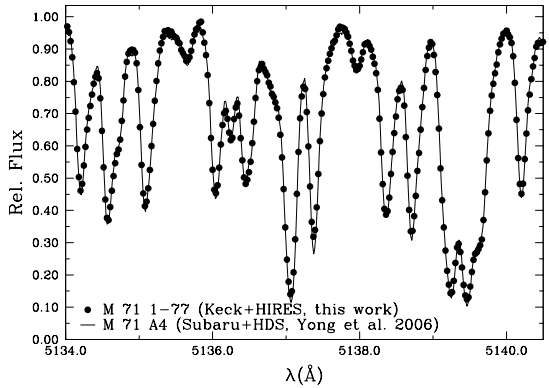}
  \caption{Comparison of M71 1-77 observed with Keck+HIRES (circles, this work)
  and M71 A4 observed with Subaru+HDS (line, Yong et al. 2006).
  Both stars have similar stellar parameters.}
\end{figure}

\section{Mg isotopes in M71 giants}
Five giants in the globular cluster M71 ([Fe/H] = -0.7, Ram\'{\i}rez et al. 2001) 
were observed with HIRES at R = 10$^5$. 
We present the analysis of three of them and 
also of M71 A4 obtained with the HDS at Subaru by Yong et al. (2006).
The latter reduced spectrum was kindly made available to us by 
D. Yong \& W. Aoki.

In Fig. 1 we compare our M71 1-77 Keck spectrum with the Subaru spectrum 
of M71 A4 (Yong et al. 2006). As it can be seen, 
even though the spectra are of two different stars (although of 
similar stellar parameters), the similarity is very impressive, 
showing that both data reductions are in excellent agreement.

The atmospheric parameters have been determined as in
Cohen et al. (2001). Iron lines were used to estimate the
microturbulence, [Fe/H] and to check the stellar parameters.
The iron lines were carefully selected in order to avoid
blends by atomic and CN lines. CN blends were visually inspected
by comparing a synthetic spectrum computed with laboratory CN lines
(e.g. Mel\'endez \& Barbuy 1999; Coelho et al. 2005) with
the high resolution visible atlas of the cool giant Arcturus
(Hinkle et al. 2000). 

Reliable laboratory oscillators strengths are not available for 
a large fraction of the FeI lines, so the lines with accurate
oscillator strengths were used to provide the zero point of
astrophysical $gf$-values. The oscillator strengths for FeII lines
are from the laboratory normalization performed by Mel\'endez et al. (2006).

A good determination of the stellar intrinsic broadening is necessary
for a reliable determination of Mg isotope ratios. The intrinsic
broadening is due to both rotation and macroturbulence 
(e.g. Hekker \& Mel\'endez 2007), but in old metal-poor stars
we expect the intrinsic broadening to be mostly due to
macroturbulence. In these cool metal-rich giants the
usual diagnostics for macroturbulence (Ni I 5115.4 \& Ti I 5145.5 \AA) 
seem blended so other lines were used for the determination of 
the macroturbulence velocity.

As for the field dwarfs, the isotope ratios in giants were determined 
from three regions, except that in our cool giants the 5134.6 \AA\ 
feature seems blended, so instead we use the 5134.3 \AA\ feature.
A $\chi^2$ fit for the 5140.2 \AA\ region is shown in Figure 2.

Our Mg isotope ratios are shown in Figure 3, where a comparison
with models (Fenner et al. 2003) is also shown. Our data favors massive stars instead of
intermediate-mass AGB stars even at the high metallicity
of M71 ([Fe/H] = -0.7)

\begin{figure}
  \includegraphics[height=.2\textheight]{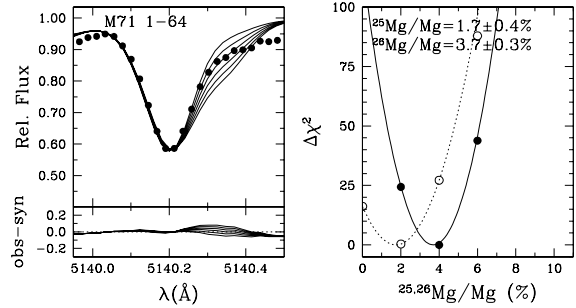}
  \caption{Fits for the 5140.2 \AA\ region in the giant M71 1-64.
Observed spectra are represented with filled circles, and synthetic
spectra with solid lines. The calculations were performed for
$^{25,26}$Mg/Mg ratios of 0, 2, ... 10 \%. The relative variation of the $\chi^2$ fits
are shown as a function of the isotopic abundance.}
\end{figure}

\begin{figure}
  \includegraphics[height=.35\textheight]{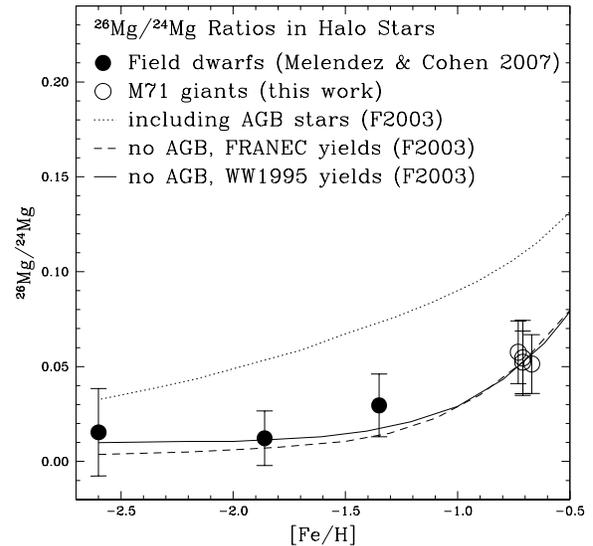}
  \caption{$^{26}$MgH/$^{24}$MgH as a function of [Fe/H] in halo dwarfs
(filled circles) and M71 giants (open circles).
Models including massive stars and intermediate-mass
AGB stars (Fenner et al. 2003 [F2003]) are also shown.
The no AGB model agrees better with the observed data.}
\end{figure}

\section{O, Na, Mg and Al in M71 giants}

We have also determined abundances of O, Na, Mg and Al in M71 giants.
The abundances were determined by both equivalent widths and
spectral synthesis. 

Unlike other clusters that show large abundance variations
(e.g. Cohen \& Mel\'endez 2005), the four giants in M71 have
essentially identical O, Na, Mg and Al abundances.
Note that the Mg isotope ratios in these four giants
is also constant within the errors (Fig. 3).
High resolution observations of a larger 
number of M71 giants will be important in order to determine
how homogeneous this cluster is.

The oxygen abundance of M71 giants seems undepleted, and consistent
with the constant [O/Fe] ratio for halo stars found by
Mel\'endez et al. (2006) and Ram\'{\i}rez, Allende Prieto \& Lambert (2007),
in the broad metallicity range $-3.2 <$ [Fe/H] $< -0.4$.


\section{Conclusions}

Our $^{26}$Mg/ $^{24}$Mg ratios in both field dwarfs and
M71 giants can be explained by massive stars (e.g. Fenner et al. 2003).
Even at the high metallicity of M71 ([Fe/H] = -0.7) there is
no need to invoke an important contribution from intermediate-mass AGB stars.

We plan to obtain more high resolution high S/N HIRES spectra of more 
field halo dwarfs and M71 giants.


\begin{theacknowledgments}
We thank D. Yong \& W. Aoki for kindly providing a Subaru spectrum
of M71 A4.  JM acknowledges partial support from the Australian Research 
Council to Martin Asplund. JGC is grateful for partial support to NSF grant  AST-0507219.
\end{theacknowledgments}



\bibliographystyle{aipproc}   

\bibliography{sample}

\IfFileExists{\jobname.bbl}{}
 {\typeout{}
  \typeout{******************************************}
  \typeout{** Please run "bibtex \jobname" to optain}
  \typeout{** the bibliography and then re-run LaTeX}
  \typeout{** twice to fix the references!}
  \typeout{******************************************}
  \typeout{}
 }

\end{document}